\documentclass[11pt]{article}
\usepackage[left=2.5cm,right=2.5cm,top=2.5cm,bottom=2.5cm]{geometry}
\usepackage{multicol}

\usepackage{amstext}
\usepackage{amsmath}
\usepackage{graphicx}
\usepackage{epstopdf}
\usepackage{multicol,caption}
\usepackage{cases}
\usepackage{amsmath,amsfonts,amssymb}
\usepackage{rotating}
\usepackage{color}
\newcommand{\red}[1]{\textcolor{black}{#1}}

\begin{document}
\pagestyle{empty}
\parindent=0mm
\parskip=0mm

\begin{center}
\large{\bfseries{Impurity seeding for suppression of the near Scrape-Off Layer heat flux \red{feature} in tokamak limited plasmas  }}\\


\normalsize{F. Nespoli$^1$, B. Labit$^1$, I. Furno$^1$, C. Theiler$^1$, U. Sheikh$^1$, \red{C.K. Tsui$^{1,2}$, J.A. Boedo$^2$, }and the TCV team$^*$}\\

\end{center}
\small{\itshape
{  $^1$Ecole Polytechnique F\'ed\'erale de Lausanne (EPFL), Swiss Plasma Center (SPC),\\
CH-1015 Lausanne, Switzerland\\
$^2$University of California-San Diego, La Jolla, California 92093, USA\\
$^*$See author list of S. Coda et al 2017 Nucl. Fusion 57 102011
\\
Email: nespolifederico@gmail.com
}}

\normalsize
\section* {Abstract}
In inboard-limited plasmas, foreseen to be used in future fusion reactors start-up and ramp down phases, the Scrape-Off Layer (SOL) exhibits two regions: the ``near'' and ``far'' SOL. The steep radial gradient of the parallel heat flux associated with the near SOL can result in excessive thermal loads onto the solid surfaces, damaging them and/or  limiting the operational space of a fusion reactor. In this \red{article}, \red{leveraging the results presented in [F. Nespoli et al., Nuclear Fusion 2017]}, we propose a  technique for the mitigation and suppression of the near SOL heat flux \red{feature} by impurity seeding. First successful experimental results from the TCV tokamak are presented and discussed.
 \\
-------------------------------------------------------------------------------------------------------------------------\\

\section{Introduction}
\begin{figure}[t]
\centering
\includegraphics[scale=0.8]{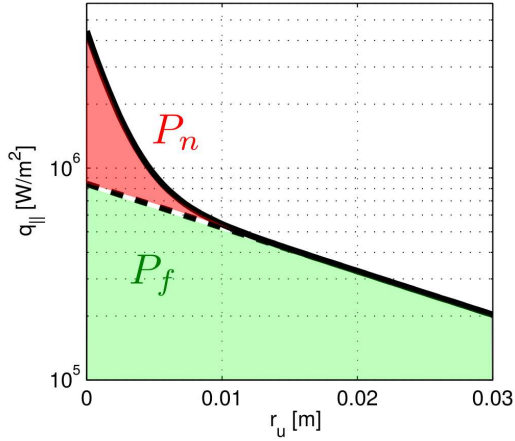}
\caption{\red{Schematics of a typical  parallel heat flux radial profile $q_{||}(r_u)$ given by  Eq. (\ref{eq:qparfit})  (solid line). The dashed line represents the heat flux associated with the far SOL, $q_f\exp(-r_u/\lambda_f)$.  The power entering the near and far SOL, $P_n$ and $P_f$ respectively,  are given by the integral of the red and green shaded areas. \label{fig:pnpf}}}
\end{figure}

Recent measurements in inboard-limited L-mode plasmas in many tokamaks \cite{Arnoux2013,Nespoli2015,Horacek2015,Stangeby2015,Marmar2015,Bak2016} with \red{both} infrared (IR) thermography and \red{reciprocating} Langmuir probes  have revealed the presence of two regions in the Scrape-Off Layer (SOL): a ``near'' SOL, extending  typically a few mm from the last closed flux surface (LCFS), characterized by a steep gradient of the parallel heat flux, and a ``far'' SOL, typically a few cm wide, featuring flatter heat flux profiles. The parallel heat flux radial profile in the SOL is then well described by a sum of two exponentials
\begin{equation}
 q_{||}(r_u)=q_{n} \exp(-r_u/\lambda_n)+q_{f} \exp(-r_u/\lambda_f)\;,\label{eq:qparfit}
\end{equation}
where $r_u$ is the upstream radial coordinate at the outer midplane, $r_u=0$ at the LCFS, $\lambda_n$, $\lambda_f$ are the parallel heat flux decay length in the near and far SOL, respectively, and $q_n$ and $q_f$ are the associated parallel heat flux magnitudes. \red{An example of a typical  parallel heat flux radial profile $q_{||}(r_u)$ described by Eq. (\ref{eq:qparfit}) is shown in Fig. \ref{fig:pnpf} with a solid line.}\\
The near SOL is responsible for the peak heat loads on the limiter, that can be a factor of 6 higher \cite{Kocan2015} with respect to \red{value expected from} the standard picture of the SOL \cite{Stangeby}, where only one decay length is assumed. Inboard limited L-mode plasmas are foreseen for future fusion reactor start-up and ramp-down phases. In ITER, \red{the beryllium (Be) tiles covering the central column will act as a limiter}. The  ITER First Wall (FW) panels ha\red{ve} been \red{recently redesigned} to handle the heat flux associated with the near SOL \cite{Kocan2015} that would otherwise exceed the Be tiles engineering constraint $q_{dep}\leq5$ MWm$^{-2}$. Still, the physics of the near SOL is not completely understood yet, and the possibility exists that the actual $\lambda_n$ in ITER will not match the one assumed \red{in Ref. \cite{Kocan2015}} for the FW panel redesign. 
 \\ \red{One can divide the power entering the SOL, $P_{SOL}$, into the contributions from the near and far SOL respectively, $P_n$ and $P_f$, such that $P_{SOL}=P_n+P_f$. A schematics of this separation is depicted in Fig. \ref{fig:pnpf}}. We define  power in the near SOL  as 
\begin{equation}
 P_{n}=4\pi R_{LCFS}\dfrac{B_{\theta,u}}{B_{\phi,u}}\int_0^\infty q_n e^{-r_u/\lambda_n}dr_u=4\pi R_{LCFS}\dfrac{B_{\theta,u}}{B_{\phi,u}}q_n\lambda_n\;,\label{eq:pn}
\end{equation}
 where $R_{LCFS}$ is the major radius of the LCFS at the outer midplane,   $B_{\theta,u}$ and $B_{\phi,u}$ are the poloidal and toroidal components, respectively, of the magnetic field at the outer midplane.  \red{Similarly, we compute the power into the far SOL as
\begin{equation}
 P_{f}=4\pi R_{LCFS}\dfrac{B_{\theta,u}}{B_{\phi,u}}q_f\lambda_f\;.\label{eq:pf}
\end{equation}}
The method proposed and tested in this \red{article} to \red{suppress} $P_n$  relies on the physics understanding gained in extensive,  dedicated  experiments performed in the TCV tokamak \cite{CodaMST}, at EPFL, Switzerland, detailed in Ref. \cite{Nespoli2017}. 
In this reference, the power in the near SOL has been shown to scale  as \red{$ P_n\propto 1/\nu$}, where $\nu=\dfrac{e n_{e}R_{0}\eta_{||}}{m_ic_{s}}$ is the normalized Spitzer resistivity, with $R_{0}$ the plasma major radius, $c_s$ the ion sound speed, $\eta_{||}$ the Spitzer resistivity \cite{Goldston}, and all quantities are evaluated at the plasma edge. Furthermore, $ P_n$ has been shown to vanish for $\nu\sim7\cdot10^{-3} $, achieved by reducing the plasma current or by increasing the density. The disappearance of the near SOL happens for values of the SOL collisionality $\nu^*_{SOL}$ corresponding to a conduction-limited  regime, being $\nu^*_{SOL}=L/\lambda_{ee}\propto n_eT_e^{\red{-}2}$ \cite{Stangeby} where $L=2\pi R_0 q_{edge}$ is the connection length and $\lambda_{ee}$ is the electron-electron collisional mean free path.
\\Based on  the dependence of the power entering the near SOL upon the normalized resistivity $ P_n\propto 1/\nu$ and its vanishing for high resistivity/collisionality,  several methods could be envisaged to suppress  the near SOL heat flux \red{feature}, or at least reduce it, and thus to prevent excessive inner wall heat loads  in a future fusion reactor (ITER, DEMO...).\\
The mitigation of the near SOL heat flux  by reducing the plasma current is not possible for a start-up scenario. Indeed,   a minimum $I_p$ is required to create a diverted configuration, which might not be low enough to prevent the formation of the near SOL. Increasing the density might not be a viable solution since  wall pumping is usually strong during the start-up phase \red{\cite{Phillips2013,Mayer2001}}, resulting in a rather  low collisionality. 
Also, the heat flux on the limiter \red{associated with the near SOL has been measured to} first increase with $n_e$ at low densities (sheath-limited regime), \red{rolling} over at intermediate densities (corresponding to the conduction limited regime) and eventually decreas\red{ing} to negligible values for high densities, if this results in a sufficient drop of temperature along the field line \cite{Nespoli2017}. Increasing the density could then  increase the heat fluxes, reaching high heat loads that could damage the FW panels.
\\In this paper, we investigate the possibility of \red{suppressing the near SOL heat flux feature} by reducing the SOL temperature, since \red{$\nu\propto T_e^{-2}$}. This  can be done, for example, by increasing the radiated power $P_{rad}$ via impurity seeding. \red{Impurity seeding, routinely employed in divertor detachment experiments, has been  extensively used in limited plasmas to both cool the edge plasma and increase confinement, e.g. in TEXTOR \cite{Samm1993,Messiaen1996,Finken2000} and JET \cite{Maddison2003}. However, even if the presence of two scale lenghts in the parallel heat flux at the limiter has been observed in these experiments  \cite{Finken2000}, the effect of impurity seeding on the near SOL has never been previously investigated. Furthermore, the possibility to use impurity seeding to completely suppress the near SOL heat flux feature has never been considered.}\\
\section{Experimental setup and and experiment overview \label{sec:setup}}
\begin{figure}[p]
\centering
\includegraphics[scale=0.8]{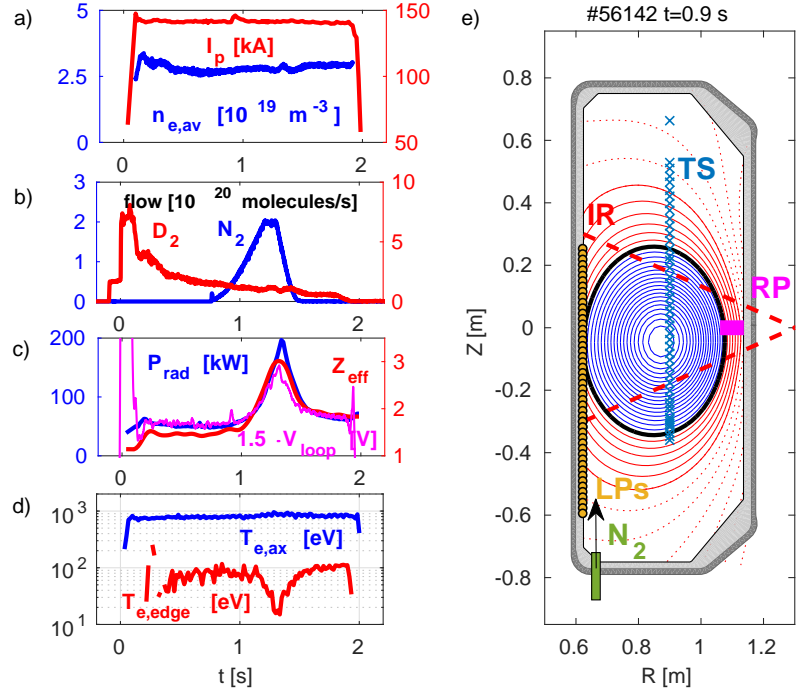}
\caption{Time traces of a)  line-averaged electron density $n_{e,av}$ (blue) and  plasma current $I_p$ (red)  b) N$_2$ \red{(blue) and D$_2$ (red)}    flow measured by the piezoelectric valve c) total radiated power $P_{rad}$  from bolometric measurements \red{(blue), the plasma effective charge $Z_{eff}$ (red) and the loop voltage $V_{loop}$ (magenta), rescaled for plotting} d) electron temperature on axis $T_{e,ax}$ (blue) and in the edge region $T_{e,edge}$ (red) from Thomson scattering measurements. e) TCV cross section together with the magnetic equilibrium reconstruction provided by LIUQE \cite{Hofmann1988}. The IR camera field of view (red dashed lines), the location of the flush mounted LPs (orange dots) \red{and of the TS measurements (blue crosses), the trajectory of the RP (magenta thick line)}, 
and the position of the valve used for N$_2$ injection (green rectangle) are also shown. \label{fig:discharge}}
\end{figure}
\begin{figure}[t]
\centering
\includegraphics[scale=0.5]{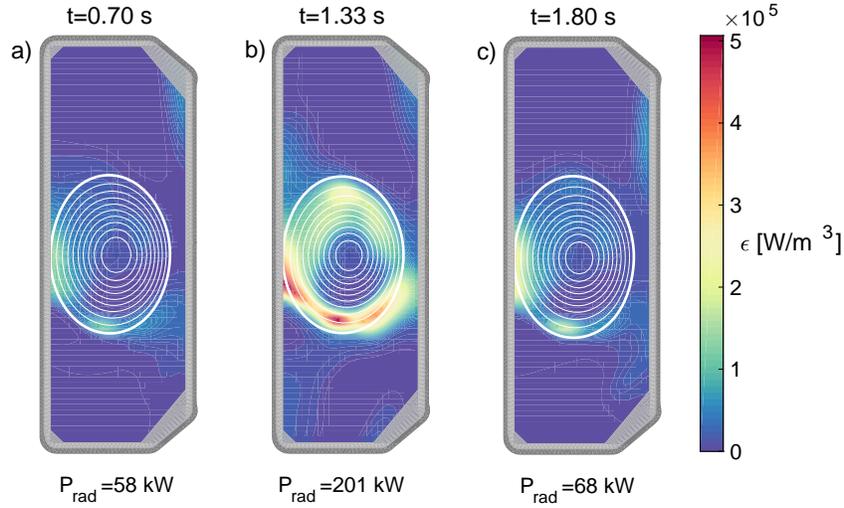}
\caption{\red{Plasma emissivity $\epsilon$ for discharge \#56142 before (a), during (b) and after (c) N$_2$ injection,  computed from the tomographic inversion of 64 gold foil bolometers measurements.} \label{fig:bolo}}
\end{figure}
\begin{figure}[!b]
\centering
\includegraphics[scale=0.8]{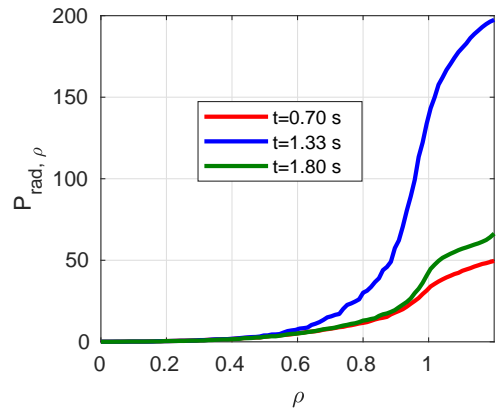}
\caption{\red{Power radiated inside each flux surface $P_{rad,\rho}$ for discharge \#56142 before (red), during (blue) and after (green) N$_2$ injection.} \label{fig:prad}}
\end{figure}
\begin{figure}[t]
\centering
\includegraphics[scale=0.8]{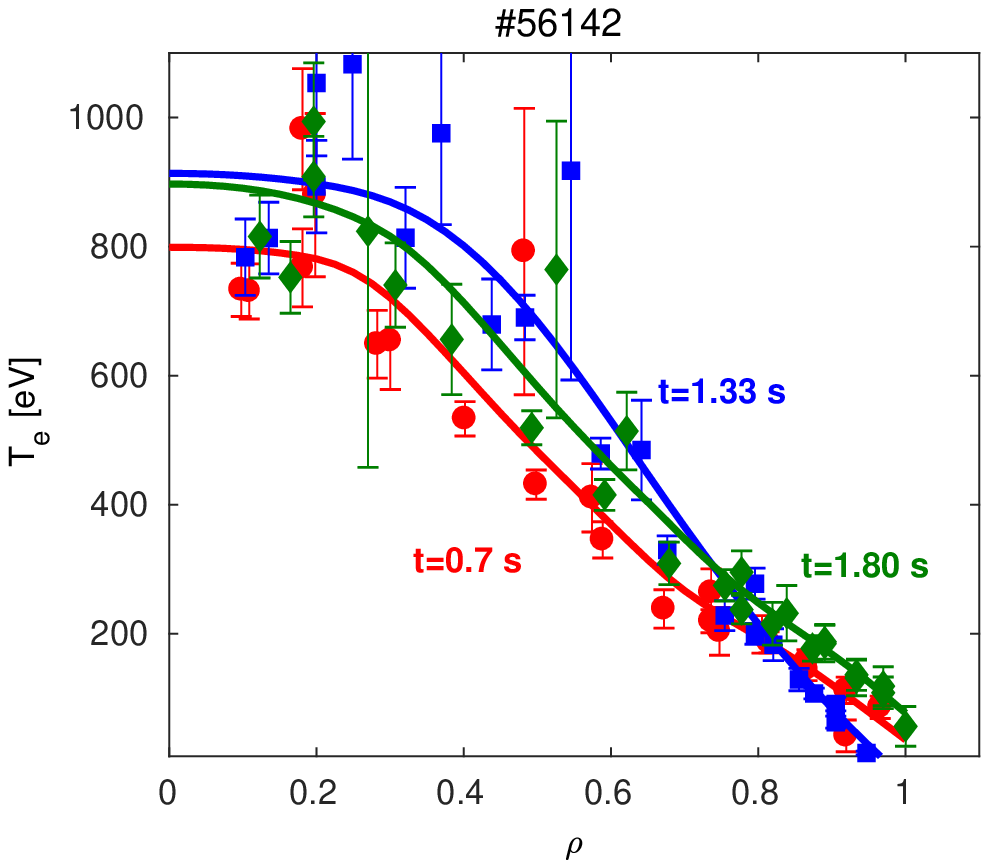}
\caption{\red{Electron temperaure radial profile $T_e(\rho)$ from Thomson scattering measurements  for discharge \#56142 before (red circles), during (blue squares) and after (green diamonds) N$_2$ injection. Smooth interpolated profiles are plotted with solid lines.} \label{fig:thomson}}
\end{figure}
To test this method, a set of dedicated experiments have been performed in  TCV, where the SOL plasma is cooled by the progressive injection of N$_2$.  An example discharge is \#56142,  which is summarized in Fig. \ref{fig:discharge}. This is an ohmically heated deuterium plasma, and the main plasma parameters are  $I_p=140$ kA,  $n_{e,av}=2.5\cdot10^{19}\;\text{m}^{-3}$, $\kappa=1.4$, $\delta=0$. While the plasma current $I_p$ and averaged density $n_{e,av}$ are kept constant  (Fig. \ref{fig:discharge}a), together with the magnetic equilibrium (Fig. \ref{fig:discharge}e reconstructed by the LIUQE code \cite{Hofmann1988}), nitrogen  (N$_2$)  is injected through a piezoelectric valve located on the TCV floor (green rectangle in Fig. \ref{fig:discharge}e). The gas flow (\red{in blue in} Fig. \ref{fig:discharge}b) is increased up to the constant level of $2\cdot10^{20}$ molecules/s, and then decreased back to zero\red{, for a  total of N$_2$  injected molecules corresponding to roughly 18\% of the injected D$_2$ molecules (in red in Fig. \ref{fig:discharge}b).} This leads to an increase  \red{of the plasma effective charge $Z_{eff}$ (in red in Fig. \ref{fig:discharge}c) by a factor 3, corresponding to an increase} of the total radiated power $P_{rad}$ (\red{in blue in} Fig. \ref{fig:discharge}c) by four times. Approximately 100 ms after the N$_2$ injection ends\red{, $Z_{eff}$ and $P_{rad}$ are decreased back to approximately 1.3 and 1.4 times their values before the N$_2$ injection, showing a slight accumulation of impurities. \red{The loop voltage $V_{loop}$ (in magenta in Fig. \ref{fig:discharge}c) exhibits a similar evolution, consistent with the changes in the plasma resistivity due to the variation of $Z_{eff}$. The plasma effective charge $Z_{eff}$ is computed by matching the plasma current using the ohmic and bootstrap current obtained from Ref. \cite{Sauter1999,Sauter2002}, using $n_e$ and $T_e$ measurements from Thomson scattering (TS), and assuming stationary state. $P_{rad}$   and the plasma emissivity $\epsilon$  are computed from the tomographic inversion of 64 gold foil bolometers measurements. Before the N$_2$ injection, $\epsilon$ is localized at the plasma contact point with the wall (Fig. \ref{fig:bolo}a). During the impurity puff (Fig. \ref{fig:bolo}b), the emissivity peaks below the contact point, consistently with the gas being injected from the bottom of the vessel. Also, $\epsilon$ is increased on the HFS part of the edge plasma, consistently with impurity penetration and with poloidal asymmetries in the impurity distribution previously observed  in other tokamaks \cite{Churchill2013}. The emissivity in the core is less affected than in the edge plasma, the level of emitted radiation remaining unchanged for $\rho\leq0.6$, where $\rho=\sqrt(1-\Psi/\Psi_{ax})$ is the normalized flux coordinate, with $\Psi$ the poloidal flux of the magnetic field and $\Psi_{ax}$ its value on the magnetic axis. The formation of a ``radiating mantle'', similarly to the TEXTOR experiments \cite{Samm1993},  is clearly shown in Fig. \ref{fig:prad}, where the the power radiated inside each flux surface $P_{rad,\rho}=\iint\epsilon(R,Z)\Theta(\rho-\rho'(R,Z))2\pi R\text{d}R\text{d}Z$ is shown, with $\Theta$ the Heaviside step function. Also, after the N$_2$ injection, $P_{rad,\rho}$ is increased, with respect to the pre-seeding situation, only for $\rho>0.9$.
 This suggests that the residual accumulation of impurities does not lower the main plasma temperature.} The increase of $P_{rad}$} in  results indeed mainly in the cooling of the edge  plasma, as shown by Thomson scattering measurements (Fig. \ref{fig:discharge}d).  The electron temperature in the edge region $T_{e,edge}\equiv\langle T_e(\rho>0.9)\rangle_\rho$ is decreased below 30 eV (in red in the figure), while the electron temperature on the magnetic axis $T_{e,ax}$ (in blue in the figure) \red{is slightly increased, resulting in the steepening of the temperature profile. After the N$_2$ injection ends, both $T_{e,edge}$ and $T_{e,ax}$ are increased 
with respect to their pre-seeding values by approximately 15\% and 60\%, respectively.  The 
steepening of the temperature profile during impurity seeding, consistent with the results from  previous seeding experiments in limited plasmas \cite{Samm1993}, and the following general increase of $T_e$, are more clearly shown in Fig. \ref{fig:thomson}, where  the radial profile of the electron temperature $T_e(\rho)$ is shown before (red), during (blue) and after (green) N$_2$ injection.
\\During the plasma discharge, } 
the temperature of the central column (CC) tiles is measured by means of IR thermography. \red{The procedure used to compute the  parallel heat flux radial profiles $q_{||}(r_u)$ is detailed in Ref. \cite{Nespoli2017}  and the main steps are summarized in the following. The heat flux deposited on the graphite tiles $q_{dep}$ is evaluated using the THEODOR code \cite{Herrmann2001}, and  is remapped onto the magnetic coordinates $(r_u,\alpha)$, where $\alpha$ is the angle between the magnetic field line and the plane tangent to the tile surface. The background and cross-field components of the heat flux, $q_{BG}$ and $q_\perp(r_u)$ are evaluated. The parallel heat flux is finally computed inverting the relationship $q_{dep}(r_u,\alpha)=q_{||}(r_u)\sin\alpha+q_{\perp}(r_u)\cos\alpha+q_{BG}$. $q_{||}(r_u)$ is computed with a time resolution of 50 ms.}
\red{The  ion saturation current $I_{sat}$ and the floating potential $V_{fl}$ are measured by a}n array of flush-mounted Langmuir probes (LPs) embedded in the limiter\red{, where every other probe is kept either  at constant biasing voltage $V=-100$ V ($I_{sat}$), either  floating ($V_{fl}$).} \red{A reciprocating Langmuir probe (RP), on loan from UCSD \cite{Boedo2009}, is located at the outer midplane of the device and can perform up to two reciprocations during a discharge.} The field of view of the IR camera,  the position of the LPs \red{and  of the RP are} shown in   Fig. \ref{fig:discharge}e.\\ 
\section{Suppression of near SOL heat flux feature and velocity shear layer at the limiter \label{sec:IRLP}}
\begin{figure}[!t]
\centering
\includegraphics[scale=0.8]{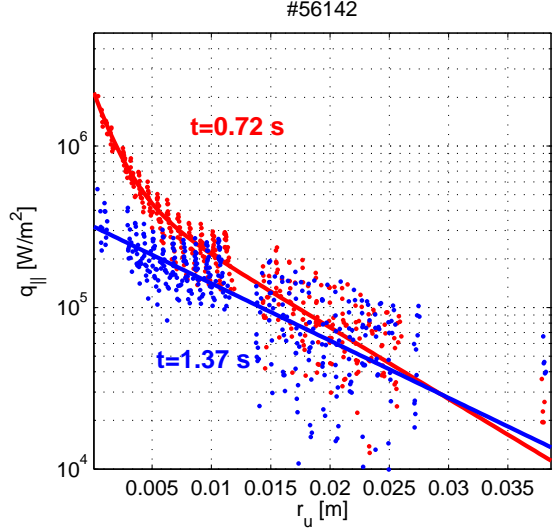}
\caption{Parallel heat flux radial profiles $q_{||}(r_u)$ before N$_2$ injection (red dots) and for $f_{rad}>70\%$ (blue dots). The fit with Eq. (\ref{eq:qparfit}) is shown with solid lines.\label{fig:IR}}
\end{figure}
\begin{figure}[!t]
\centering
\includegraphics[scale=0.8]{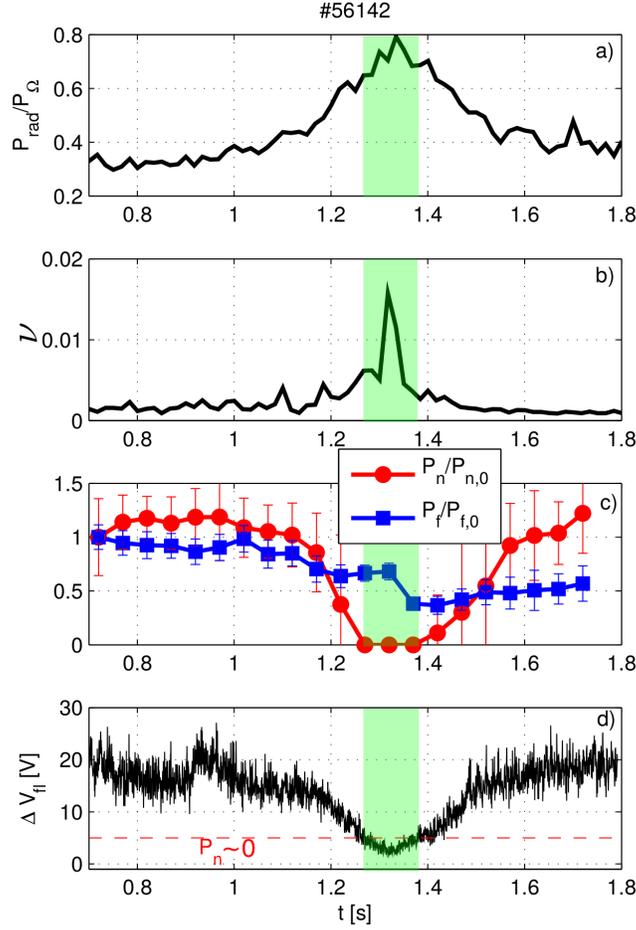}
\caption{Time traces of a)  radiated fraction  $f_{rad}=P_{rad}/P_\Omega$ b) normalized Spitzer resistivity $\nu$ c) power into the near SOL $P_n$ (red circles) and power into the far SOL $P_f$ (blue squares), normalized to their values before N$_2$ injection,  $P_{n,0}$ and $P_{f,0}$ d) drop in the floating potential $\Delta V_{fl}$ (black) compared to the value  of $\Delta V_{fl}$ for which the near SOL is observed to disappear in TCV deuterium plasmas (dashed red line). The time window for which the near SOL is suppressed is depicted with a green shadowed region.\label{fig:results}}
\end{figure}
\begin{figure}[!t]
\centering
\includegraphics[scale=0.8]{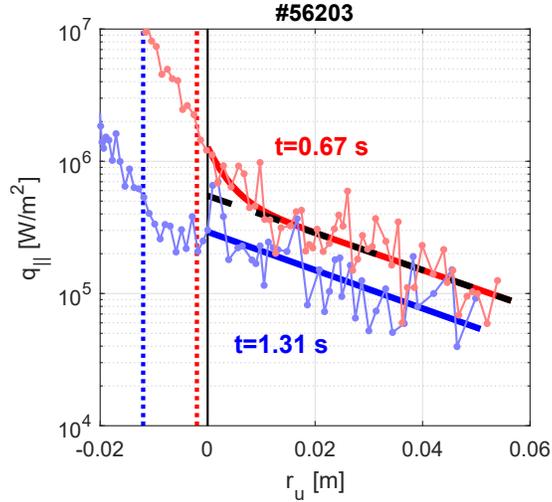}
\caption{\red{Radial profiles at the OMP of  parallel heat flux $q_{||}(r_u)$ , before N$_2$ injection (red) and for $f_{rad}>70\%$ (blue).  The fit of $q_{||}(r_u)$ with Eq. (\ref{eq:qparfit}) is shown with thick lines, while the heat flux associated with the far SOL $q_{||,f}(r_u) = q_f \exp(−r_u/\lambda_f)$ is shown with a black dashed line for the case before N$_2$ injection. The LCFS position from LIUQE is marked by a black vertical line. The LCFS location according to the method used in Ref. \cite{Tsui2017} is shown with vertical dotted lines.}\label{fig:RP}}
\end{figure}
\begin{figure}[!t]
\centering
\includegraphics[scale=0.8]{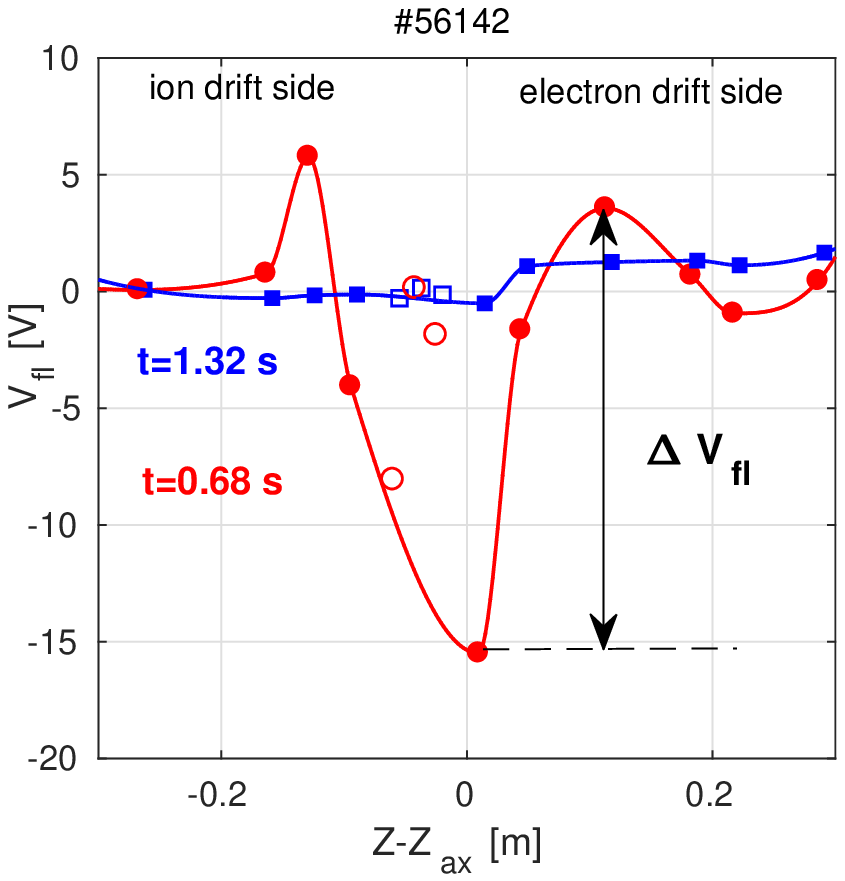}
\caption{Floating potential profile along the direction of the limiter, $Z-Z_{ax}$, with $Z_{ax}$ the vertical position of the plasma magnetic axis, before N$_2$ injection \red{(red dots)} and for $f_{rad}>70\%$ (blue squares). The profiles are interpolated with cubic splines. The drop in the floating potential $\Delta V_{fl}$ is shown. \red{Measurements from LPs} shaded by the neighboring tiles \red{are plotted with open symbols}. \label{fig:LP}}
\end{figure}
A typical \red{parallel heat flux radial} profile before the N$_2$ injection is shown in Fig. \ref{fig:IR} with red dots. Such profiles are fit to Eq. (\ref{eq:qparfit}) to determine the physical parameters $\lambda_n$, $q_n$, $\lambda_f$ and $q_f$. The result of the fit is shown in Fig. \ref{fig:IR} with a red solid line. The power into the near \red{and far} SOL, $P_n$ \red{and $P_f$,} can then be computed using Eqs. (\ref{eq:pn},\ref{eq:pf}).
The main results are exposed in Fig. \ref{fig:results} and are detailed in the following. In Fig. \ref{fig:results}a the radiated fraction $f_{rad}=P_{rad}/P_\Omega$ is shown, being 
 $P_\Omega$ the ohmic power. The cooling of the plasma results in an increase of the normalized Spitzer resistivity $\nu$ (Fig. \ref{fig:results}b)  above $7\cdot10^{-3}$, value for which the near SOL has been reported to disappear in TCV. 
The effect of the gas injection, resulting in the successful suppression of the near SOL \red{heat flux feature}, is evident from the vanishing of the power in the near SOL $P_n$ at high values of $f_{rad}$ (or $\nu$).  This is shown in Fig. \ref{fig:results}c, where
the time evolution of $P_n$ and $P_f$, normalized to their values before the N$_2$ injection $P_{n,0}$ and $P_{f,0}$, are shown  with red dots and blue squares, respectively. \red{As the N$_2$ flow is increased, t}he power in the near SOL $P_n$ \red{decreases gradually and} drops to zero \red{when} $f_{rad}\geq70\%$\red{. $P_n$} recovers its initial value  approximately 100 ms after the N$_2$ injection. A $q_{||}(r_u)$ radial profile for which $P_n=0$ is shown with blue dots in Fig. \ref{fig:IR}, together with its fit with Eq. (\ref{eq:qparfit}) (solid blue line). The power in the far SOL $P_f$ is less affected, being reduced only by  50\%. \red{Though, after the N$_2$ injection ends, $P_f$ does not fully recover to initial value. This is consistent with the residual presence of the ``raditing mantle''  in the outer edge and SOL region, as it emerges from the previously discussed bolometric measurements, resulting in a colder far SOL. }\\

For completeness of information, we remark that the increase in resistivity due to the N$_2$ injection suppresses the near SOL heat flux feature not only at the limiter, but also at the outer midplane (OMP), consistently with the results presented in Ref. \cite{Nespoli2017,Tsui2017}. The $q_{||}(r_u)$   profiles (computed from $T_e$ and $I_{sat}$ measured by the RP with the methodology detailed in \cite{Tsui2017})  are plotted in Fig. \ref{fig:RP}  for a typical discharge (\#56203), with a red (blue)  line for before (during) the N$_2$ injection. These profiles rely on the magnetic reconstruction provided by LIUQE, for which the location of the LCFS at the OMP is affected by an uncertainty of a few mm, and no corrective shift as in Ref.  \cite{Tsui2017} is applied. The application of such a corrective shift would not affect the final conclusion: similarly to the heat flux profiles measured at the limiter,  $q_{||}(r_u)$ before N$_2$ injection is well fit by a sum of two exponentials (Eq. \ref{eq:qparfit}, red thick line in Fig. \ref{fig:RP}). During the gas injection, the $q_{||}(r_u)$ profile in the SOL is sufficiently well described by a single exponential (blue solid line in Fig. \ref{fig:RP}). Though, the analysis of the poliodal asymmetries in the near SOL heat flux feature has been carried out in details in Ref.  \cite{Tsui2017}, and is beyond the objectives of this work, that aims at suppressing the near SOL heat flux feature at the limiter. \\

In the following, we discuss LPs measurements of the floating potential $V_{fl}$ at the limiter. This has a two-fold motivation: first, confirming the suppression of the near SOL as it appears from IR measurements; secondly, we propose the $V_{fl}$ measurements as a \red{trigger for an actuator} for real time monitoring of the presence of the near SOL in a start-up phase limited plasma. \\
Indeed, the presence of near SOL steep gradients is correlated with the presence of \red{local} non-ambipolar currents flowing to the limiter \cite{Dejarnac2015,Loizu2017}. These currents  cause, in TCV limited plasmas,  a drop in the floating potential radial profiles $V_{fl}(r_u)$ at the limiter, measured with LPs, that reach strong negative values as one approaches the LCFS. A $V_{fl}$ profile in the direction along the limiter before N$_2$ injection is shown with red dots in Fig. \ref{fig:LP}. The shape of this typical floating potential radial profiled has been recently explained in Ref. \cite{Loizu2017} as the result of the competition between the turbulence-driven polarization current and the poloidally asymmetric diamagnetic current, the former dominating in the near SOL. This theoretical model also predicts the progressive flattening of the $V_{fl}$ profiles with increasing SOL collisionality (i.e. resistivity $\nu$).\\
 The drop in the floating potential $\Delta V_{fl}\equiv V_{fl,max}-V_{fl,min}$, with  $V_{fl,max}$ and $V_{fl,min}$ the maximum and minimum value of $V_{fl}$, here evaluated on the electron drift-side of the limiter, can be considered as a measurement of the $\mathbf{E}\times\mathbf{B}$ shearing rate $\omega_{\mathbf{E}\times\mathbf{B}}$, being in a first approximation  $\omega_{\mathbf{E}\times\mathbf{B}}\approx\Delta V_{fl}/B\lambda_n^2$ \red{\cite{Nespoli2017}, where we approximated the gradient of the plasma potential as $\nabla V_{pl}=\nabla V_{fl} + \Lambda\nabla T_e\approx\nabla V_{fl} \approx \Delta V_{fl}/\lambda_n$. Indeed, in the near SOL,  $V_{fl}$  typically varies of $\sim20$ V over a few mm ($\lambda_n$ in the radial direction), a much stronger variation than for $\Lambda T_e$, with $\Lambda\sim3$ for deuterium plasmas}. According to a recent theoretical model \cite{Halpern2017}, the presence of a strong shear layer results in a steepening of the pressure profile, creating a separation in between near and far SOL. In Fig. \ref{fig:results}d, the time evolution of $\Delta V_{fl}$ is plotted in black, and compared with the value  $\Delta V_{fl}=5$ V, for which the power in the near SOL $P_n$ has been previously reported to vanish in TCV deuterium plasmas \cite{Nespoli2017}. As the gas injection progresses, the $\Delta V_{fl}$  is reduced, reaching values below 5 V, corresponding to a low $\omega_{\mathbf{E}\times\mathbf{B}}$, which according to the model in Ref. \cite{Halpern2017} would  be no longer sufficient to change the turbulence  and create the separation in between near and far SOL. A $V_{fl}$ profile corresponding to this phase is shown with blue squares in Fig. \ref{fig:LP}.  As the N$_2$ injection ends, $\Delta V_{fl}$ recovers its initial value and the previous shear layer is restored.\\
 \red{The $V_{fl}$ measurements from flush mounted LPs provide a reliable indication of the presence of a near SOL heat flux feature.
 These do not require an elaborated analysis, like for the IR camera, and it can produce reliable results immediately after each discharge, or even in real time, and could therefore be used as \red{a trigger for an actuator} during the start-up  and ramp-down phases. For this purpose, if one already knows the position of the plasma and the shape of the $V_{fl}$ profile at the limiter, only  measurements from two different probes per limiter side are needed. Otherwise, to correctly identify  the maximum of $V_{fl}$, at least four probes per limiter side could be needed.
We remark that this technique is somehow similar to the one used in ASDEX-U to detect detachment, where the thermoelectric current flowing to the divertor plate is measured with a single shunt, and used as a proxy of the heat flux deposited on the divertor plate \cite{Kallenbach2015}}.

\section{Implications for plasma start-up phase in a fusion reactor\label{sec:startup}}
The main drawback of our method is the temporary increase in plasma resistance, potentially increasing  the poloidal magnetic flux consumption during the start-up phase, $\Phi_{tot}$. This quantity is indeed of great importance for the design and operation of a fusion reactor, determining both the size of the central solenoid  for a steady state tokamak during its design phase \cite{Wakatsuki2017}, and  the duration of the inductive part of a plasma discharge. The total poloidal flux consumption can be decomposed as \cite{Wakatsuki2017}
$\Phi_{tot}=\Phi_{ext}+\Phi_{int}+\Phi_{res}$, where $\Phi_{ext}$ and $\Phi_{int}$ are the respectively the external and internal inductive flux consumption due to the plasma current outside and inside the LCFS, and are fixed by the plasma shape and current and by the current profile, respectively. $\Phi_{res}$ is the resistive flux consumption due to the resistive dissipation of magnetic energy, and can be expressed as \cite{Wakatsuki2017,Ejima1982,Menard2001}
\begin{equation}
\Phi_{res}=\int_{t_0}^{t1} \dfrac{dt}{I_p}\int j_\phi E_\phi dV=\int_{t_0}^{t1} dt R_\Omega I_p=\int_{t_0}^{t1}  dt V_{loop} \label{eq:phires}\;,
\end{equation} 
 where $t_0$ and $t_1$ are the beginning and ending times of the current ramp-up,  $j_\phi$ and $E_\phi$ are the toroidal plasma current density and electric field, $R_\Omega$ is the total plasma resistance and  $ V_{loop}$ is the loop voltage. The last two equalities of eq.\eqref{eq:phires} make it clear that  $\Phi_{res}$ is proportional to the time integral of the plasma resistivity (or effective charge $Z_{eff}$). The increase of $\nu$ due to the impurity injection would therefore cause an increase in $\Phi_{res}$. However, a more resistive plasma leads to a lower current penetration time. This allows a higher current ramp rate, which is limited by the rising of MHD instabilities \cite{Jardin1994}. A faster current ramp  would in turns decrease the flux consumption  $\Phi_{tot}$ allowing a decrese of $t_1$ in eq. \eqref{eq:phires}. As it was shown in Ref. \cite{Jardin1994}, the two effects would balance out leaving the total flux consumption unchanged within a few percent or less as the plasma resistivity is increased.\\
Moreover,   $\Phi_{res}$ can be reduced by the use of non-inductive current sources. Start up scenarios with reduced flux consumption have been planned using neutral beam \cite{Wakatsuki2015} and electron cyclotron \cite{Wakatsuki2017} current drive for JT60-SA and DEMO respectively. Current drive sources could then be used in a fusion reactor together with impurity injection to mitigate the heat fluxes on the limiter without increasing $\Phi_{res}$.
\\Finally, we remark that it is also possible to calibrate the level of  injected impurity, used in this work to suppress the near SOL heat flux feature, to just reduce the deposited near SOL heat flux under the  engineering constraints (5 MW/m$^2$ for ITER FW Be panels \cite{Kocan2015}) without completely suppressing it, while keeping a sufficiently low plasma resistance during the start-up phase.

\section{Conclusion \label{sec:conclusion}}
Concluding, leveraging the results from previous experiments in TCV, a  method \red{based on nitrogen seeding} to \red{suppress} the near SOL heat  flux \red{feature}   has been proposed and tested.
The results from the IR together with LPs measurements show that the near SOL heat flux \red{at the limiter} is successfully \red{suppressed} by the gas injection for $f_{rad}>70\%$, and the initial situation is \red{almost} restored after the N$_2$ injection ends, the increase of $Z_{eff}$ of $\sim30\%$ with respect to its initial value showing a slight accumulation of impurities in the main plasma. RP measurements show that the near SOL heat flux feature is successfully suppressed at the OMP as well. 
Also, we demonstrate  how   LPs  embedded in the limiter are a reliable diagnostic  to monitor the presence of a near SOL through $V_{fl}$ measurements.\\
Even though the extrapolation of  impurity transport, whose analysis is beyond the objectives of this work, from TCV to an ITER-scale fusion reactor is difficult, the modest pollution of the main plasma after the end of the gas injection might render this mitigation method suitable for a start-up phase in a fusion reactor.  Furthermore, bolometric measurements show that the injected impurity radiate mainly in the outer edge plasma and the SOL, creating a ``radiating mantle'', still present after the end of the N$_2$ injection, even with substantially lower $f_{rad}$. This has a twofold beneficial effect of both decreasing the heat flux in the far SOL, as shown from IR measurements, and to increase the main plasma temperature, as shown by Thomson scattering measurements.\\
The effect of impurity seeding on the poloidal magnetic flux consumption during the start-up phase $\Phi_{tot}$ has been discussed. The  value of $\Phi_{tot}$ could be kept unchanged even during impurity seeding by one or more of the following: i) the increase of the plasma ramp-up rate allowed by the increased plasma resistivity ii) the use of non inductive current sources iii) the tuning of the injected impurity level for keeping the deposited heat flux on the limiter below the engineering limits without totally suppressing the near SOL heat flux  feature.

\section*{Acknowledgments}

This work has been carried out within the framework of the EUROfusion Consortium and has received funding from the Euratom research and training programme 2014-2018 under grant agreement No 633053. The views and opinions expressed herein do not necessarily reflect those of the European Commission. This work was supported by the U.S. Department of Energy under Grant No. DE-SC0010529.

\bibliography{bibliography}{}

\begin{thebibliography}{10}

\bibitem{Arnoux2013}
G.~Arnoux et~al.
\newblock \textbf{Scrape-off layer properties of ITER-like limiter start-up
  plasmas in JET}.
\newblock {\em Nuclear Fusion}, \textbf{53}:073016, 2013.

\bibitem{Nespoli2015}
F.~Nespoli et~al.
\newblock \textbf{Heat loads in inboard-limited L-mode plasmas in TCV}.
\newblock {\em Journal of Nuclear Materials}, \textbf{463}:393--396, 2015.

\bibitem{Horacek2015}
J.~Horacek et~al.
\newblock \textbf{Narrow heat flux channels in the COMPASS limiter scrape-off
  layer}.
\newblock {\em Journal of Nuclear Materials}, \textbf{463}:385--388, 2015.

\bibitem{Stangeby2015}
P.C.~Stangeby et~al.
\newblock \textbf{Power deposition on the DIII-D inner wall limiter}.
\newblock {\em Journal of Nuclear Materials}, \textbf{463}:389--392, 2015.

\bibitem{Marmar2015}
E.S.~Marmar et~al.
\newblock \textbf{Alcator C-Mod: research in support of ITER and steps beyond
  }.
\newblock {\em Nuclear Fusion}, \textbf{55}:104020, 2015.

\bibitem{Bak2016}
J.G.~Bak et~al.
\newblock \textbf{Measurement of inner wall limiter SOL widths in KSTAR
  tokamak}.
\newblock {\em Nuclear Materials and Energy}, \textbf{article in press}, 2016.

\bibitem{Kocan2015}
M.~Kocan et~al.
\newblock \textbf{Impact of a narrow limiter SOL heat flux channel on the ITER
  first wall panel shaping}.
\newblock {\em Nuclear Fusion}, \textbf{55}:033019, 2015.

\bibitem{Stangeby}
P.C. Stangeby.
\newblock \textbf{The Plasma Boundary of Magnetic Fusion Devices }.
\newblock {\em Institute of Physics Publishing Bristol and Philadelph}, 2000.

\bibitem{CodaMST}
S.~Coda et~al.
\newblock \textbf{Overview of the TCV Tokamak Program: Scientific Progress and
  Facility Upgrades}.
\newblock {\em Nuclear Fusion}, \textbf{57}:102011, 2016.

\bibitem{Nespoli2017}
F.~Nespoli et~al.
\newblock \textbf{Understanding and suppressing the near Scrape-Off Layer heat
  flux feature in inboard-limited plasmas in TCV}.
\newblock {\em Nuclear Fusion}, \textbf{accepted for publication
  https://doi.org/10.1088/1741-4326/aa84e0}, 2017.

\bibitem{Goldston}
R.J. Goldston and P.H. Rutherford.
\newblock \textbf{Introduction to Plasma Physics}.
\newblock {\em Bristol: Institute of Physics Publishing}, 1997.

\bibitem{Phillips2013}
V.~Phillips et~al.
\newblock \textbf{Dynamic fuel retention and release under ITER like wall
  conditions in JET}.
\newblock {\em Journal of Nuclear Materials}, \textbf{438}:S1067--S1071, 2013.

\bibitem{Mayer2001}
M.~Mayer et~al.
\newblock \textbf{Hydrogen inventories in nuclear fusion devices}.
\newblock {\em Journal of Nuclear Materials}, \textbf{290-293}:381--388, 2001.

\bibitem{Samm1993}
U.~Samm et~al.
\newblock \textbf{Radiative edges under control by impurity fluxes}.
\newblock {\em Plasma Physics and Controlled Fusion}, \textbf{35}, 1993.

\bibitem{Messiaen1996}
A.M.~Messiaen et~al.
\newblock \textbf{High confinement and high desnity with stationary plasma
  energy and strong edge radiation in the TEXTOR-94 tokamak}.
\newblock {\em Physical Review Letters}, \textbf{77}:2487, 1996.

\bibitem{Finken2000}
K.H. Finken~T. Denner and G.~Mank.
\newblock \textbf{Thermal load distribution on the ALT-II limiter of TEXTOR-94
  during RI mode operation and during disruptions}.
\newblock {\em Nuclear Fusion}, \textbf{40}:339, 2000.

\bibitem{Maddison2003}
G.P.~Maddison et~al.
\newblock \textbf{Impurity-seeded plasma experiments on JET}.
\newblock {\em Nuclear Fusion}, \textbf{43}:49--62, 2003.

\bibitem{Hofmann1988}
F.~Hofmann and G.~Tonetti.
\newblock \textbf{Tokamak equilibrium reconstruction using Faraday rotation
  measurements}.
\newblock {\em Nuclear Fusion}, \textbf{28}:1871--1878, 1988.

\bibitem{Sauter1999}
O.~Sauter et~al.
\newblock \textbf{Neoclassical conductivity and bootstrap current formulas for
  general axisymmetric equilibria and arbitrary collisionality regime}.
\newblock {\em Physics of Plasmas}, \textbf{6}:2834, 1999.

\bibitem{Sauter2002}
O.~Sauter et~al.
\newblock \textbf{Erratum: Neoclassical conductivity and bootstrap current
  formulas for general axisymmetric equilibria and arbitrary collisionality
  regime}.
\newblock {\em Physics of Plasmas}, \textbf{9}:5140, 2002.

\bibitem{Churchill2013}
R.M.~Churchill et~al.
\newblock \textbf{In–out impurity density asymmetry in the pedestal region of
  Alcator C-Mod}.
\newblock {\em Nuclear Fusion}, \textbf{53}:122002, 2013.

\bibitem{Herrmann2001}
A.~Herrmann.
\newblock \textbf{Limitations for divertor heat flux calculations of fast
  events in tokamaks}.
\newblock {\em ECA}, \textbf{25A}:2109--2112, 2001.

\bibitem{Boedo2009}
J.A.~Boedo et~al.
\newblock \textbf{Fast scanning probe for the NSTX spherical tokamak}.
\newblock {\em The Review of Scientific Instruments}, \textbf{80}:123506,
  2009,.

\bibitem{Tsui2017}
C.~Tsui et~al.
\newblock \textbf{Poloidal asymmetry in the narrow heat flux feature in the TCV
  scrape-off layer}.
\newblock {\em Physics of Plasmas}, \textbf{24}:062508, 2017.

\bibitem{Dejarnac2015}
R.~Dejarnac et~al.
\newblock \textbf{Understanding narrow SOL power flux component in COMPASS
  limiter plasmas by use of Langmuir probes}.
\newblock {\em Journal of Nuclear Materials}, \textbf{463}:381--384, 2015.

\bibitem{Loizu2017}
J.~Loizu et~al.
\newblock \textbf{Scrape-off-layer current loops and floating potential in
  limited tokamak plasmas}.
\newblock {\em Journal of Plasma Physics}, \textbf{accepted for publication},
  2017.

\bibitem{Halpern2017}
F.~D. Halpern and P.~Ricci.
\newblock \textbf{Velocity shear, turbulent saturation, and steep plasma
  gradients in the scrape-off layer of inner-wall limited tokamaks}.
\newblock {\em Nuclear Fusion}, \textbf{57}:034001, 2017.

\bibitem{Kallenbach2015}
A.~Kallenbach et~al.
\newblock \textbf{Partial detachment of high power discharges in ASDEX
  Upgrade}.
\newblock {\em Nuclear Fusion}, \textbf{55}, 2015.

\bibitem{Wakatsuki2017}
T.~Wakatsuki et~al.
\newblock \textbf{Reduction of poloidal magnetic flux consumption during plasma
  current ramp-up in DEMO relevant plasma regimes}.
\newblock {\em Nuclear Fusion}, \textbf{57}:016015, 2017.

\bibitem{Ejima1982}
S.~Ejima et~al.
\newblock \textbf{Volt-second analysis and consumption in Doublet III plasmas}.
\newblock {\em Nuclear Fusion}, \textbf{22}:1313, 1982.

\bibitem{Menard2001}
J.E.~Menard et~al.
\newblock \textbf{Ohmic flux consumption during initial operation of the NSTX
  spherical torus}.
\newblock {\em Nuclear Fusion}, \textbf{41}:1197, 2001.

\bibitem{Jardin1994}
S.C.~Jardin et~al.
\newblock \textbf{Poloidal flux linkage requirements for the International
  Thermonuclear Experimental Reactor}.
\newblock {\em Nuclear Fusion}, \textbf{34}:1145, 1994.

\bibitem{Wakatsuki2015}
T.~Wakatsuki et~al.
\newblock \textbf{Simulation of plasma current ramp-up with reduced magnetic
  flux consumption in JT-60SA}.
\newblock {\em Plasma Physics and Fusion}, \textbf{57}:065005, 2015.

\end{thebibliography}
\bibliographystyle{unsrt} 

\end{document}